\documentclass[12pt]{article}
\pdfoutput=1
\usepackage[square,comma,numbers,sort&compress]{natbib}
\usepackage{graphicx,epstopdf,amssymb,amsfonts,amsmath,amsthm,array,
mathrsfs,amscd}
\usepackage[vcentermath]{youngtab}
\usepackage{float}
\usepackage{hyperref}
\newcommand{\pbs}[1]{\let\temp=\\#1\let\\=\temp}
\numberwithin{equation}{section}
\def\be{\begin{equation}}\def\ee{\end{equation}}
\def\cvp{\raise 2pt\hbox{,}} 
 \def\tr{\mathop{\rm tr}\nolimits}

\def\la{\lambda}

\def\plb#1#2#3{{\it Phys.\ Lett.\ }{\bf B #1} (#2) #3}
\def\npb#1#2#3{{\it Nucl.\ Phys.\ }{\bf B #1} (#2) #3}

\def\prl#1#2#3{{\it Phys.\ Rev.\ Lett.\ }{\bf #1} (#2) #3}
\def\jhep#1#2#3{{\it J. High Energy Phys.\ }{\bf #1} (#2) #3}
\def\prd#1#2#3{{\it Phys.\ Rev.\ }{\bf D #1} (#2) #3}

\def\atmp#1#2#3{{\it Adv.\ Theor.\ Math.\ Phys.\ }{\bf #1} (#2) #3}

\def\pr#1#2#3{{\it Phys.\ Rep.\ }{\bf #1} (#2) #3}
\def\jmp#1#2#3{{\it J.\ Math.\ Phys.\ }{\bf #1} (#2) #3}

\def\fortphys#1#2#3{{\it Fortsch.\ Phys.\ }{\bf #1} (#2) #3}
\def\imath#1#2#3{{\it Invent math }{\bf #1} (#2) #3}

\def\jpa#1#2#3{{\it J.\ Phys.\ }{\bf A #1} (#2) #3}
\def\ahp#1#2#3{{\it Annales Henri Poincar\'e }{\bf #1} (#2) #3}

\begin{document}
%
%
{\pagestyle{empty}
\parskip 0in
\

\vfill
\begin{center}
{\LARGE The Large $D$ Limit of Planar Diagrams}



\vspace{0.4in}

Frank F{\scshape errari}
\\
\medskip
{\it Service de Physique Th\'eorique et Math\'ematique\\
Universit\'e Libre de Bruxelles (ULB) and International Solvay Institutes\\
Campus de la Plaine, CP 231, B-1050 Bruxelles, Belgique

\smallskip

Fields, Gravity and Strings\\
Center for the Theoretical Physics of the Universe\\
Institute for Basic Sciences, Daejeon, 34047 South Korea}


\smallskip
{\tt frank.ferrari@ulb.ac.be}
\end{center}
\vfill\noindent

We show that in $\text{O}(D)$ invariant matrix theories containing a large number $D$ of complex or Hermitian matrices, one can define a $D\rightarrow\infty$ limit for which the sum over planar diagrams truncates to a tractable, yet non-trivial, sum over melon diagrams. In particular, results obtained recently in SYK and tensor models can be generalized to traditional, string-inspired matrix quantum mechanical models of black holes.

\vfill

\medskip
%
\begin{flushleft}
\today
\end{flushleft}
\newpage\pagestyle{plain}
\baselineskip 16pt
\setcounter{footnote}{0}

}


%
\section{\label{s1Sec} Introduction}

Black hole studies sit at the crossing of many fundamental issues in modern theoretical physics. At the classical level, their dynamics is governed by the Einstein field equations. Solving these equations in the full non-linear regime is essential, for instance, to describe the production of gravitational waves in black hole collisions, as recently observed \cite{gravwaves}. At the quantum level, deep issues related to the mutual consistency of unitary quantum mechanics, space-time locality and the equivalence principle arise (see e.g.\ \cite{Polrev1} and references therein for a recent review). Maybe the most fundamental and puzzling problem is to find a quantum description of the black hole interior and the associated emerging notion of time.

A fruitful approach in physics is to simplify complicated problems by using toy models and ingenious approximation schemes. This is a difficult art. The model must be simple enough to be amenable to a detailed study and yet must keep all the relevant physical features of the original problem. Over the last few years, two directions in black hole research have been pursued in this spirit. One direction is based on the use of the large space-time dimension limit in classical general relativity. The other direction is based on the study of simple quantum mechanical Hamiltonians, inspired from string theory or other. The present work originally grew from an attempt to bring together these two seemingly unrelated developments.

The idea of large $d$ in general relativity was proposed by Emparan et al.\ in \cite{Emparan}. Many important features of classical black holes physics are retained in the limit, but the analytical treatment simplifies drastically. A non-exhaustive list of interesting achievements include the computation of the quasi-normal spectrum \cite{EmparanQN}, the study of black hole instabilities \cite{EmparanINST} and of the full black hole dynamics via a membrane description \cite{EmparanDYN}.

The idea of modeling black holes with ordinary quantum mechanical Hamiltonians has a much older and rich history. It relies on the holographic correspondence, which reveals that the ordinary classical description of gravity can emerge from a dual quantum description when the number of degrees of freedom become very large. The most salient examples, which follow naturally from D-brane constructions in string theory, correspond to the large $N$ limit of the quantum mechanics of $N\times N$ matrices. The original Maldacena's proposal \cite{malda} and the BFSS matrix quantum mechanics describing the D0-brane black hole \cite{D0BH} are in this class. One may consider simpler-looking models too, since any non-trivial matrix quantum mechanics is believed to display the most essential features of quantum black holes, including unitarity loss at large $N$, the quasi-normal behaviour and chaos, see e.g.\ \cite{miscrefsBH}.\footnote{If one imposes the singlet constraint, one must use models with at least two Hermitian matrices.}

The advantage of using matrix quantum mechanics is the natural relation with string theory, which makes the bulk interpretation of the model clearer. For example, it is then always possible to introduce natural localized probes and thus, in principle, to derive the large $N$ emergent black hole geometry from first principles (see e.g.\ \cite{fer1,bere}). Recently, such models were used to derive explicitly the quasi-normal behaviour at large $N$ \cite{IP} and to discover interesting phenomena associated with the infinite redshift and the crossing of the horizon from the point of view of an external observer \cite{fer2}. It is plausible that, if full mastery could be gained on matrix quantum mechanics, the  mysteries associated with quantum black holes could be lifted.

This being said, large $N$ matrix quantum mechanical models remain a huge technical challenge, because we do not know how to perform the sum over planar diagrams in the most interesting cases. In a very recent development that has attracted a lot of attention, Kitaev and followers have shown that much more manageable models with quenched disorder display some of the crucial properties of quantum black holes \cite{Kitaevetal}, including the quasi-normal behaviour, chaos and the emergence of reparameterization invariance. This certainly came as a surprise, since these models are not related in any obvious way to string theory or quantum gravity. For this reason, their bulk space-time description, if any, is likely to be difficult or unconventional. A very interesting step in trying to improve this situation has been made by Witten in \cite{witten}.\footnote{Witten's model has been further discussed in \cite{wittenmore}.} He pointed out that the crucial structure of the Feynman graphs appearing in the SYK models is nothing but an instance of the so-called melon diagrams that are known to dominate the large $N$ limit of tensor models \cite{Gurau1}. Tensor models were invented for completely different purposes, in the ongoing quest to define $d\geq 3$ dimensional quantum gravity from the continuum limit of a sum over discretized higher dimensional geometries (see e.g.\ \cite{Gurau2} and references therein). It is gratifying that the technology developed there can find an entirely different application, in providing interesting toy models for black holes. The original so-called ``colored'' tensor model of Gurau used by Witten admits more general uncolored versions \cite{Rivasseau1} which can be further generalized along the lines of Tanasa et al.\ \cite{Tanasa} and Carrozza and Tanasa \cite{Carrozzaetal}. Witten's proposal was then extended using these works by Klebanov and Tarponolsky in \cite{klebanov}.

Tensor models are interesting, but they still look rather different from the usual matrix models we would like to study. The basic degrees of freedom in \cite{witten} are $r+1$ Majorana fermion tensors $\psi^{a_{1}\cdots a_{r}}_{A}$, $1\leq A\leq r+1$, $1\leq a_{i}\leq N$, $r\geq 3$ and global symmetry group $\text{O}(N)^{r(r+1)/2}$, whereas the model in \cite{klebanov} uses one Majorana fermion rank three tensor $\psi^{abc}$ with global symmetry $\text{O}(N)^{3}$, or simple variants. On the other hand, the basic degrees of freedom of the matrix models we want to work with are matrices $X_{\mu}$ of size $N\times N$, $(X_{\mu})^{a}_{\ b}=X^{a}_{\mu\, b}$, with $1\leq \mu\leq D$ and $1\leq a,b\leq N$. These matrices may be bosonic or fermionic (or both), Hermitian or complex. In the Hermitian case, the only global symmetry group depending on $N$ is $\text{U}(N)$, under which the matrices transform in the adjoint representation. This symmetry group must be gauged, which simply means that we restrict the Hilbert space to the sector of $\text U(N)$ invariant states. For complex matrices, the gauge group can be extended to $\text{U}(N)_{\text L}\times\text{U}(N)_{\text R}$ with a transformation law $X_{\mu}\mapsto U_{\text L}X_{\mu}U_{\text R}^{-1}$. This looks much more like a tensor model, see e.g.\ \cite{Bonzom}. One is interested in the limit $N\rightarrow\infty$, which selects planar diagrams, for a fixed number of matrices $D$.

In the string theoretic D-brane picture, the number $D$ of \emph{bosonic} matrices is usually related to the number of space dimensions transverse to the D-branes. Ordinary quantum mechanics are naturally associated to D-particles, with $d-1$ transverse dimensions. It is thus natural to identify $d=D+1$. More generally, if we deal with D$p$-branes, associated with matrix quantum field theories in $p+1$ worldvolume dimensions, then we identify $d=D+p+1$. In all cases, the matrix theory is endowed with a global rotation symmetry $\text{O}(D)$, which is not gauged, under which the bosonic matrices $X_{\mu}$ transform as a vector. This set-up is realized both in the cases of the $\text{AdS}_{5}$ Schwarzschild black hole \cite{WittenBHAdS} and of the D0-brane black hole \cite{D0BH}.

This is where the idea \cite{Emparan} of using the limit of large space-time dimension $d$ in general relativity comes in. It gives us the motivation to look into the limit $D\rightarrow\infty$ in the matrix model. This idea is at the basis of the present research \cite{project}. Now that we have the motivation, we shall not try to make the relation with the results of Emparan et al.\ very precise; this is left for future investigations. In particular, even if they are philosophically similar, we do not claim that our large $D$ limit is the same as Emparan's. For example, our models are more naturally related to $\text{AdS}_{p+2}\times\text{S}^{D-1}$ bulk space-times rather than $\text{AdS}_{d}$.

We are going to concentrate on the study of the matrix model Feynman diagrams in the limit $D\rightarrow\infty$. Our main result is to show that this limit can be defined in such way that the sum over planar diagrams truncates to a non-trivial sum over melon diagrams. \emph{An absolutely crucial point is that the limits $N\rightarrow\infty$ and $D\rightarrow\infty$ do not commute. One must first take $N\rightarrow\infty$ and second $D\rightarrow\infty$.} In other words, one starts by considering the sum over planar diagrams and then takes the large $D$ limit of this sum. This yields a well-defined expansion in powers of $1/\sqrt{D}$. Note that this ordering of the limits is consistent with the intuition from Emparan et al., where the large $d$ limit is considered within general relativity, i.e.\ \emph{after} the classical limit of gravity has been taken.

The plan of the rest of the paper is as follows. In Section \ref{s2Sec}, we describe precisely our findings. In Section \ref{s3Sec}, we prove the main result. A more comprehensive account will appear in a separate paper \cite{Fernew}. In Section \ref{s4Sec}, we conclude and provide a few suggestions for future directions of research.

\section{\label{s2Sec} Description of the results}

\subsection{\label{ModelsSec} Models}

\begin{figure}
\centerline{\includegraphics[width=6in]{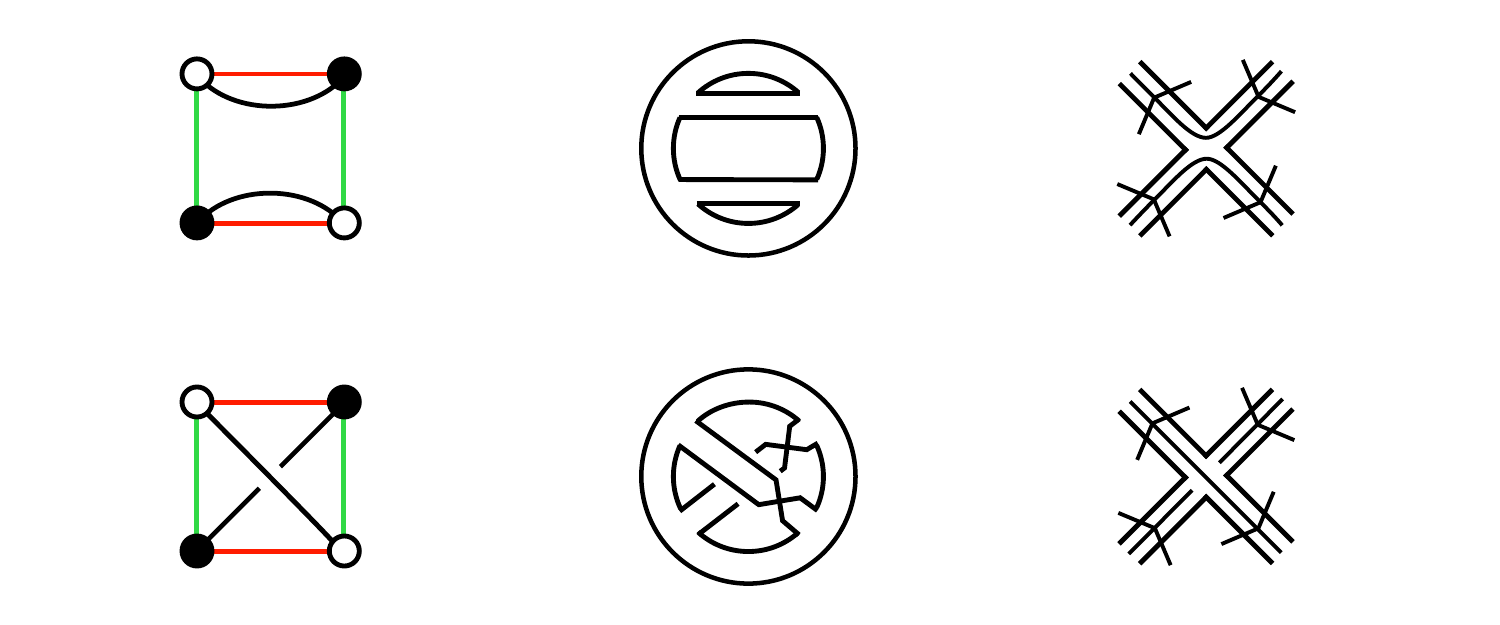}}
\caption{From left to right, colored, ribbon and stranded graphs associated to the quartic interaction terms $\tr (X_{\mu}X_{\mu}^{\dagger}X_{\nu}X_{\nu}^{\dagger})$ (up) and $\tr (X_{\mu}X_{\nu}^{\dagger}X_{\mu}X_{\nu}^{\dagger})$ (down). The r-graphs have genus zero and one-half respectively. We have chosen to orient the ribbons in the stranded graph, corresponding to the case of complex matrices.\label{fig1}}
\end{figure}

Only the general features of the Feynman diagrams are relevant for our results to hold. They apply to $(p+1)$-dimensional matrix field theories for any $p\geq -1$, but for concreteness we choose the case $p=0$ of matrix quantum mechanics. We consider Lagrangians 
\be\label{typicalH} L =ND\Bigl( \tr\bigl(\dot X_{\mu}^{\dagger}\dot X_{\mu}+m^{2}X_{\mu}^{\dagger}X_{\mu}\bigr) - \sum_{B}t_{B}I_{B}(X)\Bigr)\, ,\ee
where the $t_{B}$s are coupling constants and the $I_{B}$s are single-trace interaction terms of the form
\be\label{IBdef1} I_{B} = \tr\bigl( X_{\mu_{1}}X_{\mu_{2}}^{\dagger}X_{\mu_{3}}X_{\mu_{4}}^{\dagger}X_{\mu_{5}}\cdots X_{\mu_{2s}}^{\dagger}\bigr)\ee
for which the indices $\mu_{i}$ are identified pairwise and summed over. Each term \eqref{IBdef1} has a convenient colored graph (c-graph) representation, with lines of colors green, red and black associated with the indices $a$, $b$ and $\mu$ in $X^{a}_{\mu\, b}$ and two types of vertices, unfilled or filled, associated with each $X$ and $X^{\dagger}$ appearing in \eqref{IBdef1}. The vertices form polygons and are joined by lines according to the index contractions in \eqref{IBdef1}. If one travels along the polygon clockwise, green and red lines always join filled to unfilled and unfilled to filled vertices respectively. In particular, green and red lines always join vertices of different types: we say that they respect the bipartite structure of the graph. The black lines, on the other hand, may violate this bipartite structure. These c-graphs can be drawned on two-dimensional surfaces, which yields a standard ribbon graph (r-graph), by choosing a cyclic ordering of the colored lines around each vertices, e.g.\ (green, red, black) clockwise around unfilled vertices and anticlockwise around filled vertices. Moreover, we decide that the ribbons associated with lines joining vertices of different types are untwisted, whereas the ribbons associated with lines (necessarily black) joining vertices of the same type are twisted. This choice ensures that the faces of the (not necessarily orientable) r-graph so obtained are in one-to-one correspondence with the cycles made of lines of alternating colors in the colored graph.\footnote{Such cycles are called faces of the c-graph.} The genus $g(B)$ of the interaction $B$ is defined to be the genus of the associated r-graph. On top of the c- and r- graph representations, an interaction vertex also has the familiar stranded (s-graph) representation, which is the usual ribbon vertex of matrix models supplemented with a thread for the index $\mu$. The two independent quartic interaction terms are displayed in the three possible representations on Fig.\ \ref{fig1}.

For complex matrices, the Lagrangian \eqref{typicalH} has a $\text{U}(N)^{2}\times\text{O}(D)$ symmetry. In principle, the unitary symmetry should be gauged. However, the effect of the gauging is subleading in the large $N$ and large $D$ limits we consider. For example, the gauging yields corrections of order $N^{2}$ to the free energy, because the dimension of the gauge group is of order $N^{2}$. These corrections are negligible compared to the leading $N^{2}D$ terms.\footnote{This is unlike the case of tensor models with $\text{U}(N)^{2}\times\text{O}(N)$, which also appear naturally \cite{Tanasa,klebanov}. In the tensor context, the full symmetry group is gauged \cite{klebanovgauging}.} At subleading orders, the gauging will of course have an effect. We do not implement it explicitly, because this can be done straightforwardly without changing the main points of our discussion. 

Note that instead of bosonic matrices, one could also and very similarly consider models with fermionic matrices with obvious modifications of the Lagrangian. The Feynman graphs would be the same. 

For Hermitian matrices, the reality condition reduces the symmetry down to $\text{U}(N)\times\text{O}(D)$. The interaction terms are still written as \eqref{IBdef1} and thus can still be represented as a c-graph. However, there is an ambiguity in the c-graph coming from $X_{\mu}=X_{\mu}^{\dagger}$: c-graphs obtained from each other by swapping the filled and unfilled vertices and the green and red colors represent the same term.\footnote{This swap ambiguity is generically twofold, but some symmetric graphs, including the one associated with the $\tr (X_{\mu}X_{\nu}X_{\mu}X_{\nu})$ interaction, may be untouched by this transformation.} The genus of the r-graph is independent of the ambiguity. The s-graph simply loses the orientation of the propagators compared to the case of complex matrices.

\subsection{\label{StandardScalingResSec} The standard scaling}

To define the large $N$ and $D$ limits, we need to specify how the couplings in \eqref{typicalH} scale with $N$ and $D$. It is natural to investigate first the simplest case for which the $t_{B}$ are held fixed in the limits.\footnote{Even though we shall see that this is not the most relevant choice. This standard scaling has also been discussed in \cite{japanese}; see also \cite{italians}.} This choice corresponds to the familiar scalings both in matrix models and in vector models, if we see our collection of matrices $X_{\mu}$ as the $D$ entries of a $\text{O}(D)$ vector.\footnote{For a nice review on vector models, see e.g.\ \cite{vectorrev}.} This standard scaling can be shown to possess the following properties, which are valid either in the complex $X_{\mu}\not = X_{\mu}^{\dagger}$ or in the Hermitian $X_{\mu}=X_{\mu}^{\dagger}$ cases.

\noindent\emph{Prop.\ 1 (commutativity of the limits)}: the large $N$ and large $D$ limits in the standard scaling commute. The free energy has an expansion of the form
\be\label{Fexp1} F=\sum_{(g,L)\in\mathbb N^{2}}f_{g,L}N^{2-2g}D^{1-L}\ee
for some $N$- and $D$-independent coefficients $f_{g,L}$. Similar expansions hold for correlation functions.\smallskip\\
\noindent\emph{Prop.\ 2 (non-renormalization)}: only interaction terms having $g(B)=0$ contribute to the leading order $N^{2}D$ (and generalizations of this statement to any genus order $N^{2-2g}$ exist).\smallskip\\
\noindent\emph{Prop.\ 3 (non-renormalization)}: for the models in which the interaction terms are of the form $\tr (X_{\mu}X_{\mu}^{\dagger})^{s}$, vacuum diagrams of genus $g$ are of order $D^{1-g}$ or lower (and similar statements hold for correlation functions). In other words, at order $D^{1-L}$ in the large $D$ expansion, only diagrams of genus zero to $L$ can contribute and the coefficients $F_{g,L}$ are all zero for $g>L$.

These results are interesting because they show that the large $D$ and large $N$ limits are intertwined with one another. However, the standard scaling eliminates too many diagrams and the results are too simple to capture all the qualitative physical aspects of the usual large $N$ limit at fixed $D$. The physics is more akin to the ordinary vector models than to the matrix models. At the technical level, we can use the familar auxiliary field method to straightforwardly solve the models. The correlation functions computed in this way do not exhibit black hole behaviour; there is no unitary loss nor chaos.

\subsection{\label{NewScalingResSec} The new scaling}

We now propose a new, much more interesting, large $D$ scaling.\footnote{This new scaling was inspired in a crucial way by the work of Carrozza et al.\ on the $\text{O}(N)^{3}$ random tensor model \cite{Carrozzaetal}; see also \cite{Tanasa}.} Instead of fixing $t_{B}$, we fix
\be\label{newscaling} \lambda_{B}=D^{-g(B)}t_{B}\, ,\ee
where $g(B)$ is the genus of the r-graph representation of the interaction associated with the coupling $t_{B}$. This reproduces the standard 't~Hooft's large $N$ scaling, but is very different from the standard vector model large $D$ scaling as soon as interactions with $g(B)>0$ are included. The couplings $t_{B}$ diverge for such interactions and thus many more Feynman diagrams are kept in the limit. 

One can prove the following result, \emph{valid both in the complex and the Hermitian cases:}\smallskip\\
\noindent\emph{Prop.\ 4}: the sum over planar diagrams has a well-defined large $D$ expansion at fixed $\lambda_{B}$ in powers of $1/\sqrt{D}$. For example, if we denote by $N^{2}F_{0}$ the planar free energy, we have
\be\label{F0exp} F_{0}=\sum_{\ell\in\mathbb N}F_{0,\ell}D^{1-\ell/2}\ee
for some $N$- and $D$-independent coefficients $F_{0,\ell}$.\smallskip\\
This theorem is the main result of our work. It can be refined to all genera in the case of complex matrices. For example, the genus $g$ free energy $N^{2-2g}F_{g}$ of the complex matrix models has a large $D$ expansion of the form
\be\label{Fgexpnew} F_{g}=\sum_{\ell\in\mathbb N}F_{g,\ell}D^{1+g-\ell/2}\, .\ee
The highest power $1+g$ of $D$ at fixed genus is the optimal universal upper bound, but better bounds may be valid if only some particular vertices are included.\footnote{For example, if the interaction terms are all of the form $\tr (X_{\mu}X_{\mu}^{\dagger})^{s}$, then $\ell\geq 4g$ as a consequence of Prop.\ 3 in Sec.\ \ref{StandardScalingResSec}.} Similar bounds may still be valid for (traceless) Hermitian matrices, but we have no proof in genera $g>0$.\footnote{See \cite{Hermitiantrace,Fermult} for recent developments.} What is easy to check explicitly is that the power of $D$ in graphs of arbitrary genus is not bounded from above. This implies in particular that one cannot take the large $D$ limit with the new scaling \eqref{newscaling} at fixed $N$. \emph{The limits make sense only if one performs $N\rightarrow\infty$ first and $D\rightarrow\infty$ second.}

\begin{figure}
\centerline{\includegraphics[width=6in]{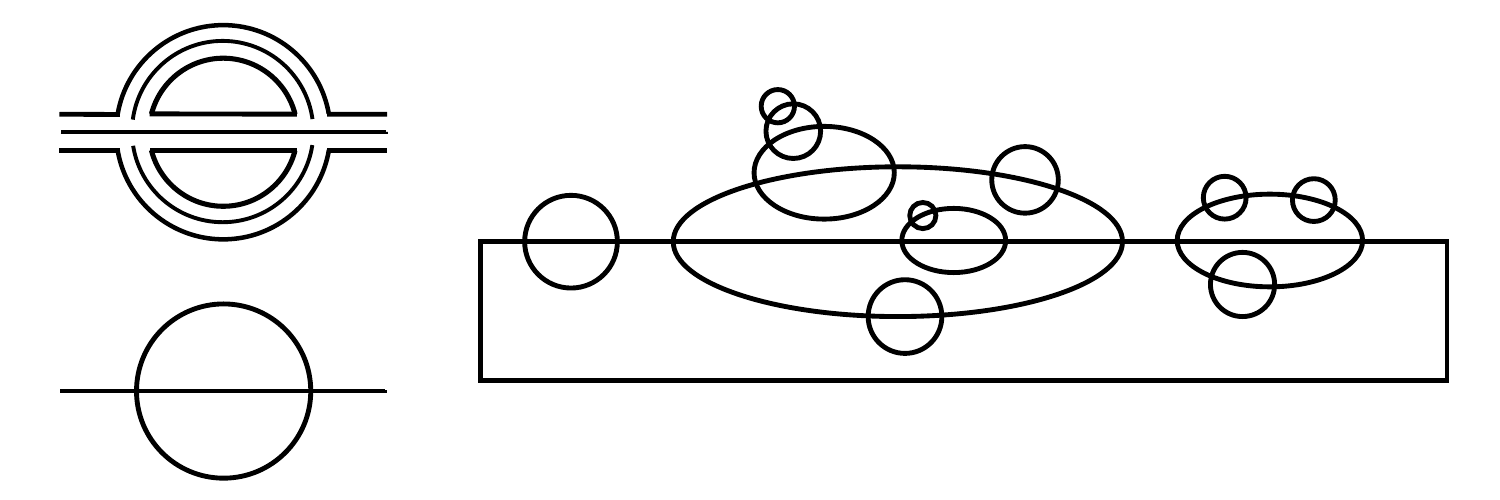}}
\caption{The basic two-loop diagram from which melons are built in the $\lambda\sqrt{D}\tr (X_{\mu}X_{\nu}X_{\mu}X_{\nu})$ model (upper-left), its stylized representation (lower left) and a typical stylized melon diagram (right). All the diagrams contributing at leading order $N^{2}D$ are of this type.\label{fig2}}
\end{figure}

The leading graphs that contribute at order $N^{2}D$ are generalized melons of the kind encountered in tensor models and SYK models. They can be built recursively. For example, if we include only the $\lambda\sqrt{D}\tr (X_{\mu}X_{\nu}X_{\mu}X_{\nu})$ interaction, the melon diagrams are built by iteratively replacing propagators by a basic two-loop diagram, see Fig.\ \ref{fig2}. Note that a bosonic model with only the $\tr (X_{\mu}X_{\nu}X_{\mu}X_{\nu})$ term is not stable, but a stable model is obtained for example by adding a term $\tr (X_{\mu}X_{\mu})^{6}$ to the action. This does not change the physics nor the math in any drastic way.\footnote{See \cite{Fermult} for examples of stable models containing bosons. In particular, it is shown in \cite{Fermult} that our new large $D$ scaling is compatible with supersymmetry. Supersymmetric models are of course automatically stable.} The recursive structure of melons imply that one can always write down closed Schwinger-Dyson equations for them. In this sense, they are under full analytic control. From these equations, one can derive the quasi-normal behavior, chaos, time reparameterization, etc., in genuine matrix models, along the lines of previous works on SYK models \cite{Kitaevetal}.

\noindent\emph{Remark on the interaction term $\tr [X_{\mu},X_{\nu}]^{2}$}: in several models derived from string theory, in particular in the D0-brane matrix quantum mechanics, the interaction term 
\be\label{Lint} L_{\text{int}} = N D \la \tr [X_{\mu},X_{\nu}]^{2} = 2 ND \la\tr \bigl(X_{\mu}X_{\nu}X_{\mu}X_{\nu} - X_{\mu}X_{\mu}X_{\nu}X_{\nu}\bigr)\ee
plays a privileged role. The reason is that it is automatically produced by dimensional reduction of the $\tr [A_{M},A_{N}]^{2}$ term in gauge theory. The structure of \eqref{Lint} is incompatible with the new large $D$ scaling \eqref{newscaling}, since the two contributions $\tr X_{\mu}X_{\mu}X_{\nu}X_{\nu}$ and $\tr X_{\mu}X_{\nu}X_{\mu}X_{\nu}$ scale differently at large $D$, see Fig.\ \ref{fig1}. The best we can do is to introduce two couplings $\la_{1}$ and $\la_{2}$ and study
\be\label{Lint2} L_{\text{int}}= 2 ND \tr \bigl(\sqrt{D}\la_{1}\, X_{\mu}X_{\nu}X_{\mu}X_{\nu} - \la_{2}\, X_{\mu}X_{\mu}X_{\nu}X_{\nu}\bigr)\ee
instead of \eqref{Lint}.

This is an unfortunate situation. Superficially, it may seem that the standard scaling, which is clearly compatible with the structure of the commutator squared, is more suitable to study \eqref{Lint}. But this is naive: Prop.\ 2 of Sec.\ \ref{StandardScalingResSec} actually implies that the term $\tr X_{\mu}X_{\nu}X_{\mu}X_{\nu}$ does not contribute at all at leading order! On the contrary, the new scaling produces a model for which both terms in \eqref{Lint} contribute at leading order. This will better capture some of the interesting physics associated with the commutator squared interaction, in particular the black-hole like properties.

\section{\label{s3Sec} The proof of the main result}

We focus on the Prop.\ 4 of Sec.\ \ref{NewScalingResSec}, which is our main result. Prop.\ 1 and 2 in Sec.\ \ref{StandardScalingResSec} are simple. Prop.\ 3 requires a non-trivial proof which will be given elsewhere \cite{Fernew}. Let us simply mention that in the case of complex matrix models, it can be shown to follow from a simple generalization of Lemma 7 in the penultimate reference in \cite{Gurau1}. For Hermitian matrix models, the tools of tensor models do not apply and one must use a different strategy. The idea is to reformulate the original model in terms of the usual auxiliary fields used in vector models. In this new formulation, we are dealing with a two-matrix model with single-trace and double-trace interactions. Prop.\ 3 follows from an analysis of the topology of the resulting Feynman diagrams.

\subsection{\label{CpxeSec} The case of complex matrices}

Let us first prove Prop.\ 4 in the case of complex matrices. The model is then $\text{U}(N)^{2}\times\text{O}(D)$ invariant and tensor model techniques can be used.\footnote{The literature has focused on $\text{U}(N)^{r}$ or $\text{O}(N)^{3}$ invariant models, but our case can be analysed along the same lines. In particular, most of the ideas in \cite{Carrozzaetal} can be adapted straightforwardly.} 

The Feynman diagrams\footnote{We focus on connected vacuum diagrams, but the analysis can be easily generalized to correlation functions as well.} have both standard s-graph and c-graph representations. The c-graph is built from the s-graph as follows. The vertices of the s-graph become 3-colored graphs $B_{a}$, $1\leq a\leq v$, as explained in \ref{ModelsSec}. The propagators become a new sort of line, say violet, that join the vertices of the c-graph. The violet lines must respect the bipartite structure of the c-graph. This is the  counterpart of the fact that the s-graph propagators for complex matrices are oriented. We get in this way a 4-colored graph representation of any Feynman diagram, with all lines except the black respecting the bipartite structure.

A c-graph \emph{face} of colors $(i,j)$, $i\not = j$, is defined to be a cycle of the c-graph made of lines of alternating colors $i$ and $j$. A crucial property of the 4-colored graphs is that their faces (violet, green) and (violet, red) on the one hand and (violet, black) on the other hand, are in one-to-one correspondence with the faces (closed loops) of the s-graph made of strands associated with $\text{U}(N)$ indices on the one hand and with $\text{O}(D)$ indices on the other hand. For a given Feynman diagram, if we denote by $f$, $\varphi$, $v$ and $p$ the numbers of $\text{U}(N)$ faces, $\text{O}(D)$ faces, vertices and propagators in the s-graph  and by $F_{ij}$ and $V$ the number of $(i,j)$-faces\footnote{For example, $F_{\text{vg}}$ is the number of (violet, green) faces, etc.} and vertices in the c-graph, then
\be\label{relationstoc} f=F_{\text{vg}} + F_{\text{vr}}\, ,\quad \varphi = F_{\text{vb}}\, . \ee
Moreover, if $v_{2s}$ is the number of $2s$-valent vertices in the s-graph, one has
\be\label{relationstoc2} V = 2\sum_{s} s v_{2s}= 2p\, .\ee
The $N$ and $D$ dependence of a Feynman graph derived from the Lagrangian \eqref{typicalH} in the scaling \eqref{newscaling} is given by
\be\label{Feynmangraph} N^{2-2g}D^{1+g-\ell/2}\, ,\ee
where $g=1+\frac{1}{2}(p-v-f)$ is the genus of the standard matrix model fat graph and
\be\label{ellformula} \ell =4-3 v + 3 p - f-2 \varphi-2 \sum_{a=1}^{v}g(B_{a})\, .\ee

We now use a strandard trick in tensor model technology. If we erase from a 4-colored graph all the lines of a given color $i$, we get a set of connected 3-colored graphs $B^{(i)}_{a}$, $1\leq a\leq B^{(i)}$, where $B^{(i)}$ is the number of connected components. To these 3-colored graphs, we can associate a ribbon graph, exactly as explained for the c-graphs representing the interactions in Sec.\ \ref{ModelsSec}. We thus get a genus $g(B^{(i)}_{a})$ for each connected component and we define
\be\label{sumg} g_{i}=\sum_{a} g(B^{(i)}_{a})\, .\ee
For example, if $i=\text{v}$, the $B^{(\text{v})}_{a}=B_{a}$ are the s-graph vertices, $B^{(\text{v})}=v$ and $g_{\text v}$ is the sum on the right-hand side of \eqref{ellformula}. If $i=\text{b}$ then there is only one connected component, $B^{(\text{b})}=1$ and it is easy to check that its genus $g_{\text b}$ coincides with the genus $g$ of the matrix model fat graph. 

It is also useful to note that this procedure of erasing lines of certain colors can be pursued. If we erase both the lines of colors $i$ and $j$ from the original Feynman c-graph, we obtain a set of 2-colored connected graphs $B^{(ij)}_{a}$, $1\leq a\leq B^{(ij)}$, which are nothing but the faces of the original graph in the complementary colors. For example, $B^{(\text{gr})}=F_{\text{vb}}$, etc.

These notions being introduced, it is a simple matter of careful but straightforward face counting to show that $\ell$ defined in \eqref{ellformula} can be expressed as
\be\label{ellformula2} \frac{\ell}{2} = g_{\text g} + g_{\text r} +\bigl(B^{(\text{vg})} - B^{(\text v)} - B^{(\text g)} + 1\bigr) + \bigl(
B^{(\text{vr})} - B^{(\text v)} - B^{(\text r)} + 1\bigr)\, .\ee

Let us now make a small diversion.\footnote{I would like to thank Tatsuo Azeyanagi for a discussion on this point.} We consider an arbitrary connected graph $\mathscr B$ made of vertices and lines joining the vertices. Let i and j be two disjoint subsets of the set of lines of $\mathscr B$ and denote by $\mathscr{B}^{(\text i)}$, $\mathscr{B}^{(\text j)}$ and $\mathscr{B}^{(\text{ij})}$ the graphs obtained from $\mathscr B$ after erasing lines in i, j, and both i and j. Denote by $B^{(\text i)}$, $B^{(\text j)}$ and $B^{(\text{ij})}$ the numbers of connected components of these graphs. Then the following inequality holds,
\be\label{contopo} B^{(\text{ij})}\geq B^{(\text i)} + B^{(\text j)} -1\, .\ee
This is proven by noting that, if we add the lines in j to $\mathscr{B}^{(\text{j})}$, we get $\mathscr B$, which is connected. The lines in j must thus connect all the $B^{(\text{j})}$ connected components of $\mathscr{B}^{(\text{j})}$ together. The very same lines thus also connect at least $B^{(\text{j})}$ connected components of $\mathscr{B}^{(\text{ij})}$ together. If we add the lines j to $\mathscr{G}^{(\text{ij})}$, the number of connected components $G^{(\text i)}$ we are left with is thus at most $G^{(\text{ij})}-G^{(\text{j})}+1$, which proves \eqref{contopo}.

The four terms in the right-hand side of \eqref{ellformula2} are thus all positive: the first two terms are positive because they correspond to sums of genera of surfaces whereas the last two terms are positive because of the general inequality \eqref{contopo}. Using \eqref{Feynmangraph}, we find that for a given genus $g$, there is an upper bound $1+g$ for the power of $D$ that can appear in a Feynman diagram. This shows that there is a well-defined large $D$ expansion at each genus. Moreover, \eqref{ellformula} or \eqref{ellformula2} show that $\ell$ is an integer. Because $g_{\text g}$ and $g_{\text r}$ can be half-integers (recall from Sec.\ \ref{ModelsSec} that the ribbon graphs made out of c-graphs containing black lines may be non-orientable), we see that the large $D$ expansion parameter is $1/\sqrt{D}$. The proof is complete.

Finally, let us also note that all the four terms in \eqref{ellformula2} must vanish for the graphs contributing at leading order. Following \cite{Carrozzaetal}, one can then prove that these leading graphs are melons of the form depicted in Fig.\ \ref{fig2}.

\subsection{\label{HermitianSec} The case of Hermitian matrices}

Superficially, the problem with Hermitian matrices is much harder. The symmetry is reduced down to $\text{U}(N)\times\text{O}(D)$, we get many more Feynman diagrams and the technology developed for tensor models is a priori useless. However, let us try to repeat the arguments of the previous subsection, identify where they fail and then try to find a way out. As we shall see, we will succeed, at least in the most interesting case of the planar diagrams. The proof is not very difficult, but we provide more details than in the previous sections, because it may seem very surprising that results which have the flavor of tensor models could still be relevant for ordinary Hermitian matrix models.

A first difference with complex matrices is that the Feynman graphs now have many possible c-graph representations. This ``swap'' ambiguity comes from the possibility to exchange vertex types and green and red colors in the representation of the s-graph vertices, as explained in Sec.\ \ref{ModelsSec}. Let us assume that a choice has been made and each s-graph vertex is represented by a 3-colored graph $B_{a}$, $1\leq a\leq v$. The Feynman graph is then built by joining the vertices of all the $B_{a}$s with violet lines associated with the propagators of the s-graph. A second difference with the case of complex matrices then arises: the violet lines do not need to respect the bipartite structure of the graph. This is the counterpart of the fact that the s-graph propagators for Hermitian matrices are \emph{not} oriented.

In itself, having lines that do not respect the bipartite structure is not a problem. Black lines already had this property in the case of complex matrices. However, the difficulty with non-bipartite violet lines is that we loose the crucial correspondence between the faces of the s-graph made of strands associated with $\text{U}(N)$ indices and the (violet, green) and (violet, red) faces of the c-graph. The new correct rule is that the s-graph $\text{U}(N)$ faces correspond to cycles of the c-graph composed of an even number of lines of alternating colors (violet, green) and (violet, red), such that: if the violet line joins vertices of different types, then the color, green or red, before and after the violet line must be the same; if the violet line joins vertices of the same type, then the color before and after the violet line must be different, yielding sequences green-violet-red.

\begin{figure}
\centerline{\includegraphics[width=6in]{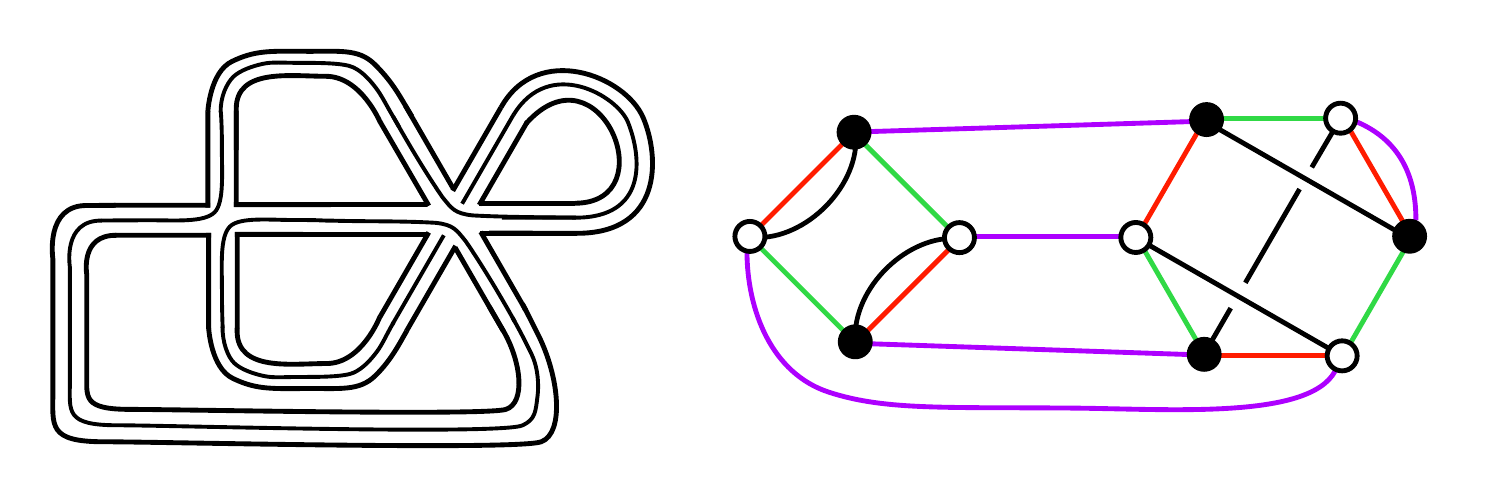}}
\caption{A planar four-loop Feynman diagram of order $N^{2}D^{-1}$ in the Hermitian matrix model in the s-graph (left) and c-graph (right) representations. The diagram contains two vertices $\tr (X_{\mu}X_{\mu}X_{\nu}X_{\nu})$ and $\tr (X_{\mu}X_{\nu}X_{\rho}X_{\nu}X_{\mu}X_{\rho})$ of genera zero and one respectively. The number of $\text{U}(N)$ faces $f=5$ does not match $F_{\text{vg}}+F_{\text{vr}}=3$.\label{fig3}}
\end{figure}

This is illustrated on Fig.\ \ref{fig3} on a simple planar four-loop diagram. The stranded graph has $f=5$, $\varphi=1$, $v=2$ and $g=0$. The formulas \eqref{Feynmangraph} and \eqref{ellformula}, which are equally valid in the complex and Hermitian cases, show that the diagram is of order $N^{2}D^{-1}$, with $\ell=4$. The colored graph has $F_{\text{vg}}=1$, $F_{\text{vr}}=2$ and $F_{\text{vb}}=1$. The first relation in \eqref{relationstoc} is violated. 

\begin{figure}
\centerline{\includegraphics[width=6in]{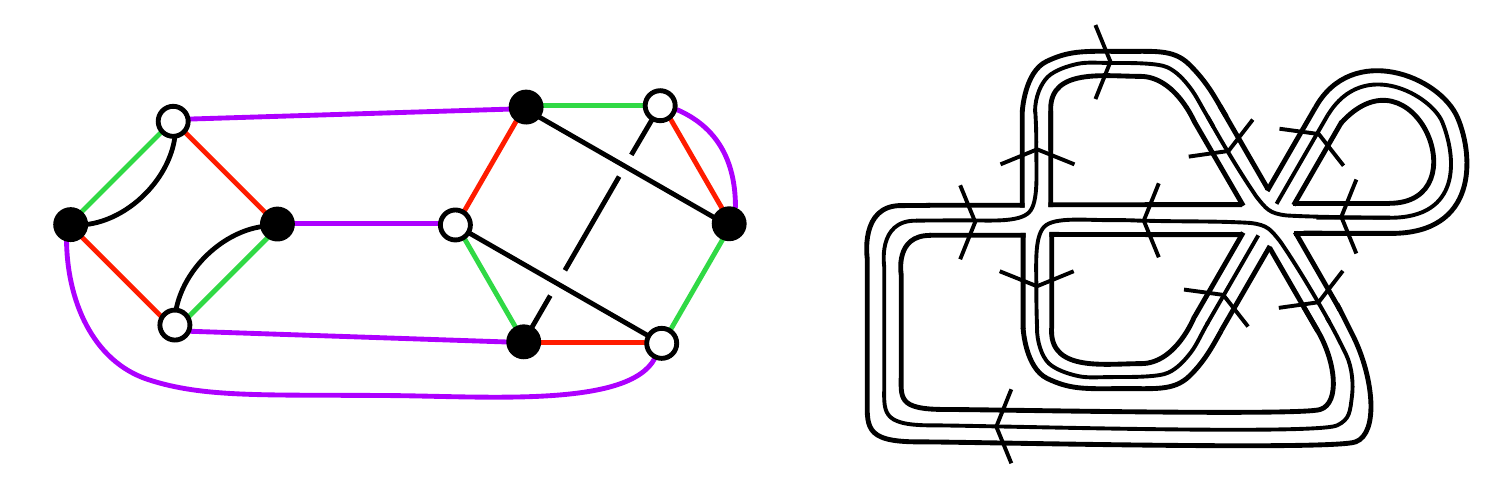}}
\caption{Another c-graph representation (left) of the Feynman graph depicted on Fig.\ \ref{fig3}, for which the violet lines respect the bipartite structure and thus for which $F_{\text{vg}}+F_{\text{vr}}=f=5$. Equivalently, one can orient the propagators of the Hermitian model s-graph to make it indistinguishable from a complex model graph (right).\label{fig4}}
\end{figure}

This lack of a direct relation between the natural notions of faces in the s-graph and c-graph representations imply that the reasoning made for complex matrices will not work here. Fortunately, there is one possible way out. If the violet lines do not respect the bipartite structure on one c-graph representation of the Feynman diagram, we may try to improve the situation by using another c-graph representation, using the swap ambiguity mentioned previously. For example, if we swap the c-graph of the quadrivalent s-vertex in the graph of Fig.\ \ref{fig3}, we get a new c-graph representation of the Feynman graph, depicted on Fig.\ \ref{fig4}, for which the violet lines respect the bipartite structure! For such a graph, the analysis of Sec.\ \ref{CpxeSec} can be repeated.

Can we do this for all Feynman graphs? It is very easy to find counterexamples at genera $g\geq 1$. However, to prove Prop.\ 4, we need to consider only \emph{planar} Feynman diagrams. And we are now going to see that indeed, for planar diagrams, one can always find one (actually two) c-graph representation for which the violet lines respect the bipartite structure. This is equivalent to the fact that it is always possible to choose  an orientation of the propagators of any planar Hermitian Feynman s-graph that makes it indistinguishable from a planar complex Feynman s-graph, see Fig.\ \ref{fig4}. One can then repeat the arguments of the previous subsection and conclude that Prop.\ 4 of Sec.\ \ref{NewScalingResSec} is valid for the Hermitian matrix models.

\begin{figure}
\centerline{\includegraphics[width=6in]{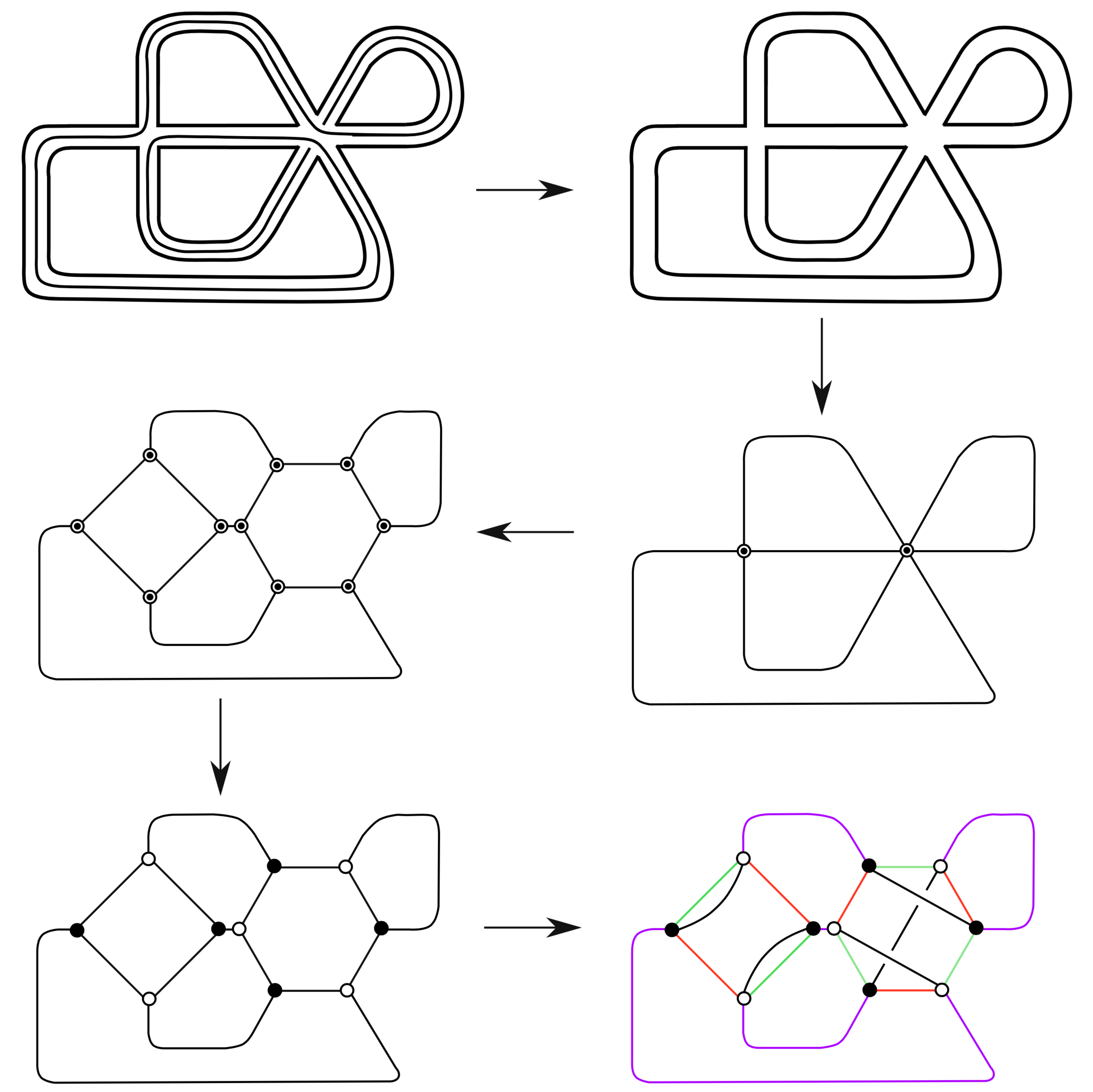}}
\caption{Steps from a stranded planar Feynman graph for the Hermitian matrix model to a c-graph whose violet lines respect the bipartite structure.\label{fig5}}
\end{figure}

Let us thus consider an arbitrary planar Feynman diagram in our Hermitian matrix models. We are going to build, step by step, a c-graph representation with the desired properties, starting from the stranded graph. The procedure is illustrated on an example in Fig.\ \ref{fig5}. 

\noindent\emph{Step 1}: we erase the $\text{O}(D)$ strands from the s-graph. This yields the usual ribbon 't~Hooft's graph which, by hypothesis, is planar.\\
\emph{Step 2}: we replace the ribbons by ordinary lines. Planarity is simply equivalent to the fact that the resulting lines do not cross.\\
\emph{Step 3}: we replace each vertex of valence $s$ by $s$ trivalent vertices arranged on a polygon. Each new vertex has two lines attached joining the adjacent vertices along the polygon and a third line pointing outwardly of the polygon. We call the resulting graph the ``skeleton.'' It is automatically planar. It matches a simplified version of a c-graph representing the Feynman diagram we started with, in which the black lines are erased and the types of the vertices (filled or unfilled) and the colors of the remaining lines (green or red) are forgotten.

The faces of the skeleton are of two types. First, we have the faces of the original 't~Hooft's graph. Because the old vertices are now polygons, each propagator line making such a face is now systematically followed by a polygon edge. This automatically yields faces made of an even number of lines. Second, we get one new polygonal face for each old vertex. Since the valency of the old vertices is even, these new faces are also made of an even number of lines. In conclusion, the skeleton is a planar graph whose faces are all made of an even number of lines.

But such planar graphs are well-known to be bipartite: it is always possible to separate the set of vertices in two classes, ``filled'' and ``unfilled,'' in such a way that all the lines join filled to unfilled vertices only. If we consider the dual squeleton graph, which is obtained in the standard way by exchanging faces and vertices, we get a planar graph whose vertices all have even valency. The result is then equivalent to saying that the faces of this dual graphs can be colored with only two colors in such a way that two adjacent faces never have the same color, a fact known as Kempe's two-color theorem.\footnote{For a general planar graph, with vertices of arbitrary valency, a very famous theorem states that no more than four colors are needed in general.} We can thus continue our construction of the c-graph.

\noindent\emph{Step 4}: endow the skeleton graph with a bipartite structure (there is an irrelevant twofold ambiguity at this step, which amounts to swapping the types of all the vertices of the graph).\\
\emph{Step 5}: reinstate the black lines. This is done unambiguously by matching with the structure of the original stranded graph.\\
\emph{Step 6}: color the remaining lines in green, red and violet to obtain the desired c-graph. This can be done unambiguously. The violet lines correspond to the ribbons (propagators) of the original stranded graph. The lines on the polygons are colored in green and red by respecting the rule explained in Sec.\ \ref{ModelsSec}: if one travels clockwise around a polygon, green lines join filled to unfilled vertices whereas red lines join unfilled to filled vertices. This completes the proof.

\section{\label{s4Sec} Conclusion}

In a groundbreaking work more than 40 years ago, 't~Hooft showed that theories made of $N\times N$ matrices have a well-defined large $N$ expansion \cite{tHooftplanar}. The set of Feynman graphs can be partitioned according to the genus of the surface on which the graphs can be drawn without line crossings and diagrams of genus $g$ are proportional to $N^{2-2g}$. The main interest in this expansion is that the leading order is believed to capture most of, if not all, the important non-perturbative physics of the models under consideration. This explains why seeking methods to sum over planar diagrams has remained a central topic in theoretical physics up to the present day. The original aim was to 
find an approximate solution to the theory of strong interactions, but through string theory and the holographic correspondence planar diagrams have found many more potential applications, including in quantum black hole physics.

We have shown in this work that in a large class of interesting $\text{O}(D)$ invariant matrix models, the sum over planar diagrams can itself be expanded at large $D$. As explained in Sec.\ \ref{NewScalingResSec}, the interesting large $D$ scaling is unlike the standard scaling used in vector models. The set of planar diagrams is partitioned according to the index $\ell$ defined in Eq.\ \eqref{ellformula} or equivalently Eq.\ \eqref{ellformula2}. Planar diagrams of index $\ell$ are proportional to $D^{1-\ell/2}$. The truly remarkable point is that the leading order in this new expansion seems to be able to capture some of the most interesting non-perturbative features of the sum over planar diagrams. Equally remarkable is that the $\ell=0$ diagrams, called generalized melons, can be summed over analytically: closed-form Schwinger-Dyson equations that fix the sums over melons unambiguously can be written down. This seems to open many new opportunities to study non-perturbative physics in matrix models, particularly in matrix quantum mechanics. 

It is fascinating to try to uncover some of the mysteries of quantum black holes using these techniques. Compared to the models studied over the last year \cite{Kitaevetal, witten, wittenmore, klebanov}, the advantage is that we are dealing here with ordinary matrix models which have a much more direct string theoretic interpretation. It seems plausible that bulk duals could be explicitly constructed. A possible relation with the work of Emparan et al.\ \cite{Emparan}, or a suitable generalization thereof, which was instrumental in motivating us to study the large $D$ limit, would be very interesting to investigate and, if valid, could greatly help in understanding the bulk physics.

Apart from the direct applications to black holes, there are many other obvious directions of research to pursue. For example, it is natural to try to study the large $D$ limit of the Hermitian matrix models beyond the planar diagrams. The relation to tensor models is then seemingly completely lost, but we believe that the limit could still make sense. The fact that the standard tensor model techniques might not be essential is suggested by our Prop.\ 3 in Sec.\ \ref{StandardScalingResSec}, which can be proven for all genera in the Hermitian case. Maybe a useful idea will be to reformulate the models in terms of auxiliary fields via the usual Hubbard-Stratonovich transformation. In general, this reformulation produces a new kind of random tensor model involving the tensor variable
\be\label{Auxvar} T^{ab}_{cd}=X^{a}_{\mu\, c}X^{b}_{\mu\, b}\ee
with non-standard interaction terms. These non-standard interactions do not prevent the tensor model to have a well-defined large $N$ expansion, since it is equivalent to the original matrix model.\footnote{Note that this idea is similar to the idea which leads from colored to uncolored tensor models, when one constructs the uncolored models as truncations of the theory obtained from the colored model by integrating out all but one tensor variable.} It may be fruitful to look at the large $D$ limit in this framework too. More generally, we feel that the class of tensor-like theories that admits interesting limits is probably much larger than what has been studied up to now, see \cite{Hermitiantrace, Fermult} for recent developments.

Finally, it is hard to resist mentioning possible applications to QCD. This is natural, since we claim to have a new powerful way to truncate the sum over planar diagrams. However, many difficulties, in relation with gauge invariance and renormalizability, prevent a direct applications of our ideas to Yang-Mills models. This is unlike the case of black holes, where the applications seem to be around the corner.

\noindent\emph{Note added in proof:} the results of the present paper have been generalised recently to matrix-tensor models of any rank and orthogonal and/or unitary symmetry groups \cite{FRV}. See also \cite{Fermult}.

\subsection*{Acknowledgments}

I would like to thank Tatsuo Azeyanagi and Paolo Gregori for interesting discussions. I would also like to thank Soo-Jong Rey for inviting me to the new Center for Fields, Gravity and Strings in South Korea and for providing a greatly stimulating scientific atmosphere there.

This research is supported in part by the Belgian Fonds National de la Recherche Scientifique FNRS (convention IISN 4.4503.15 and CDR grant J.0088.15 ``Quantum Models of Black Holes'') and the F\'ed\'eration Wallonie-Bruxelles (Advanced ARC project ``Holography, Gauge Theories and Quantum Gravity'').

%
%
%

%

%

%
\end{document}